\documentclass[useAMS,usenatbib,fleqn,twocolumn,resetfootnote]{aastex7}
\usepackage{xcolor}
\usepackage{amsmath}
\usepackage{float}
\usepackage{verbatim}
\usepackage{xspace}
\usepackage{graphicx}

\def\sgra{Sgr~A$^*$\xspace}
\def\m87{M87$^*$\xspace}

\begin{document}

\title{Observational Properties of Near-Maximally Spinning Supermassive Black Holes}

\author[orcid=0000-0001-8217-6787,sname='Thomas']{Tegan A. Thomas}
\affiliation{University of Virginia, Department of Astronomy,530 McCormick Road, Charlottesville, VA 22904, USA}
\affiliation{Washington University in St. Louis, Department of Physics,1 Brookings Drive, St. Louis, MO 63130, USA}
\email[show]{thomas.astrophysics@gmail.com}  

\author[orcid=0000-0001-5287-0452,sname='Ricarte']{Angelo Ricarte} 
\affiliation{Center for Astrophysics $\vert$ Harvard \& Smithsonian, 60 Garden Street, Cambridge, MA 02138, USA}
\affiliation{Black Hole Initiative at Harvard University, 20 Garden Street, Cambridge, MA 02138, USA}
\email{angelo.ricarte@cfa.harvard.edu}

\author[orcid=0000-0002-0393-7734,sname=Prather]{Cora Prather}
\affiliation{Black Hole Initiative at Harvard University, 20 Garden Street, Cambridge, MA 02138, USA}
\email{cprather@fas.harvard.edu}

\author[orcid=0000-0002-2858-9481,sname=Cho]{Hyerin Cho}
\affiliation{Center for Astrophysics $\vert$ Harvard \& Smithsonian, 60 Garden Street, Cambridge, MA 02138, USA}
\affiliation{Black Hole Initiative at Harvard University, 20 Garden Street, Cambridge, MA 02138, USA}
\email{hyerin.cho@cfa.harvard.edu}

\begin{abstract}
Black holes described by the Kerr metric can have a theoretical maximum dimensionless spin parameter of $a_\bullet = 1$, but several effects may limit the maximum spin parameter in astrophysical systems.  We perform general relativistic magnetohydrodynamics simulations of accretion flows around black holes with $a_\bullet = 0.9375$ and $a_\bullet = 0.998$, each corresponding to a proposed astrophysical limit in the literature.  We then perform full polarized general relativistic ray-tracing to produce astrophysical movies of these simulations, as can be spatially resolved by the Event Horizon Telescope (EHT) and its extensions.  Although many properties of black holes and accretion flows evolve rapidly as $a_\bullet \to 1$, we find that our $a_\bullet=0.9375$ and $a_\bullet=0.998$ simulations are remarkably similar, both in terms of their GRMHD fluid properties and their full-Stokes, time-variable images.  This suggests that previous work using simulations with $a_\bullet \approx 0.9375$ may be representative of models with $a_\bullet \gtrsim 0.9375$ in most practical cases.  Our calculations suggest that shape and size constraints on the photon ring, enabled by extensions of the EHT into space by missions such as the Black Hole Explorer (BHEX) may be the only practical way to distinguish between models with different spin parameters as $a\to 1$.
\end{abstract}


\section{Introduction} 

Supermassive Black Holes (SMBHs) can be found at the centers of most massive galaxies, with which they are believed to co-evolve over cosmic time \citep[e.g.,][]{Kormendy_Ho_13}. SMBHs grow predominantly via the accretion of gas supplied by their host galaxies, and in turn impart Active Galactic Nucleus (AGN) feedback on their surroundings in the form of radiation, winds, and jets \citep{Heckman_Best_14}.  It has recently become possible to produce resolved images of SMBHs via the Event Horizon Telescope (EHT) \citep{EHT_1_19,EHT_1_22}, allowing the nearest SMBHs to be utilized as laboratories to test general relativity and the metric describing black holes in the strong curvature regime.  

Since charge is expected to be efficiently discharged \citep{Blandford_Znajek_77}, astrophysical SMBHs are believed to be well-described by the \citet{Kerr_63} metric, whose only parameters are mass and spin, which we will denote as $M_\bullet$ and $a_\bullet$ respectively.  SMBH masses can be directly measured by observing their gravitational pull on gas and stars in their vicinity \citep[e.g.,][]{Saglia_etal_16,EHT_6_19,Gravity_22}.  SMBH spins, on the other hand, are much less easily estimated.  A little over 50 SMBH spins have been measured using the ``X-ray reflection spectroscopy'' method \citep{Reynolds_21,Bambi_etal_21}.  Here, a well-sampled X-ray spectrum is fit with a model containing a disk that truncates at the innermost stable circular orbit (ISCO), which moves inward as the spin increases \citep{Bardeen_70}.  One of the most notable features is the Doppler shift of the 6.4 keV Fe K$\alpha$ line, which broadens asymmetrically as spin increases \citep[e.g.,][]{Gates_Hadar_24}.  Such measurements are model-dependent and can only be obtained for the brightest objects with thin disks, but so far result in a spin distribution that peaks at maximum spin and includes no retrograde objects (wherein the disk and SMBH angular momentum vectors are anti-aligned) \citep{Piotrowska_etal_2024,Ricarte_etal_25,Mallick_etal_25}.

Spin is also imprinted on resolved images, and is expected to become more faithfully recoverable as image fidelity improves.  Moving forward, the next-generation Event Horizon Telescope (ngEHT) will improve upon the EHT by adding more radio dishes to the Very Long Baseline Interferometric (VLBI) array and observing at 3 frequencies simultaneously \citep{Johnson_etal_23}. This will result in better resolution and spatial frequency coverage for observations of M87* and Sgr A* \citep{Johnson_etal_23}. Additionally, the proposed Black Hole Explorer (BHEX) mission would further expand the EHT into space, vastly improving our resolution to the order of 5$\mu$as \citep{Johnson_etal_24}. With BHEX, additional sources besides Sgr A* and M87* will become observationally accessible, further increasing the importance of having methods with which we may constrain black hole parameters such as inclination or magnetic field strength \citep{Zhang_etal_25}.  By spatially resolving the photon ring, the image of photons marginally bound to the SMBH, BHEX will be capable of model-independent spin measurements of \m87 and \sgra with $\approx$10\% precision.  For less well-resolved SMBHs, the morphology of the linear polarized image is a promising probe of spin in simulations \citep{Palumbo_Wong_Prather_20,Qiu_etal_23,Emami_etal_23,Ricarte_etal_23}.  BHEX offers a new opportunity to measure SMBH spins for a new population of objects: SMBHs with very low specific accretion rates rather than very high accessible through reflection methods.

Apart from being a key feature of the space-time, a SMBH's spin lies at the center of its accretion and feedback processes.  A SMBH can gain or lose spin via accretion and SMBH-SMBH mergers, depending on the orientation of incoming material, and may also lose spin to powering a jet \citep[see][for a comprehensive exploration]{Ricarte_etal_25}. Specifically, for a classical thin disk, the radiative efficiency is strongly spin-dependent, roughly 5\% for $a_\bullet=0$ and 42\% for $a_\bullet=1$ \citep{Novikov_Thorne_73}.  For a jet powered via electromagnetic spin extraction, the efficiency scales approximately as $a_\bullet^2$ \citep{Blandford_Znajek_77}, even beyond $>100$\% as $a_\bullet \to 1$ \citep{Tchekhovskoy_Narayan_McKinney_10}.  The steep behavior of both of these quantities as $a_\bullet \to 1$ makes the maximum achievable value of $a_\bullet$ an astrophysically interesting problem.  \citet{Thorne_74} argued that the negative torque from a thin disk's radiation may cap a SMBH's spin at $a_\bullet=0.998$.  Meanwhile, changes to the disk dynamics and spin energy extraction via relativistic jets may limit the spin to even lower values \citep{Gammie_04,Narayan_etal_22,Ricarte_Narayan_Curd_23,Lowell_etal_24}.

This motivates our present study to perform 3D numerical simulations of highly spinning SMBHs, one model with $a_\bullet=0.9375$ and one model with $a_\bullet = 0.998$ to quantify any observable differences accessible to EHT and its extensions.  We first perform 3D General Relativistic Magnetohydrodynamics (GRMHD) simulations with a fixed spacetime metric initialized with magnetized tori of plasma using the code KHARMA.  Next, we perform polarized radiative transfer calculations using the General Relativistic Ray-tracing (GRRT) code IPOLE.  Section \ref{methods} describes the methods used. In section \ref{results} we present and analyze our preliminary results, and we discuss the implications for potential EHT observations of maximally spinning black holes in section \ref{conclusion}.

\section{Methodology} \label{methods}

We follow the Patoka pipeline to produce GRMHD fluid simulations, as well as their resultant GRRT-derived images \citep{Wong_etal_22}.  In this work, we compare two different spin values: $a_\bullet=0.9375$, a limit proposed by \cite{Gammie_04}, and $a_\bullet=0.998$, a limit proposed by \cite{Thorne_74}. 

\subsection{GRMHD Simulations}

In order to model the plasma around black holes we use the General Relativistic Magneto-Hydrodynamic (GRMHD) code known as  {\sc KHARMA} \citep{Prather_24}.  We initialize each simulation with a \citet{Fishbone_Moncrief_76} torus with an innermost radius of $r_{in}=20\,r_g$, where $r_g =GM_\bullet c^{-2} $, and a pressure maximum at $r_{max}=41\, r_g$ with an angular momentum vector parallel to that of the BH.  The plasma is initialized with a plasma $\beta$ of 100, known to produce the ``Magnetically Arrested Disk'' (MAD) state \citep{Bisnovatyi_Ruzmaikin_1974,Narayan_Igumenshchev_Abramowicz_03,Igumenshchev_etal_2003,Tchekhovskoy_Narayan_McKinney_11}. We assume an adiabatic index of $\hat{\gamma}$ of $\frac{13}{9}$, appropriate for a pure hydrogen plasma with relativistic electrons and sub-relativistic ions \citep{Shapiro_73}.  We run our GRMHD simulations to $t \approx 30,000 \,t_g$, where $t_g=GM_\bullet c^{-3}$.  We analyze only $t\geq 15,000\, t_g$, within which we verify inflow equilibrium to at least 30$r_g$. 

In order to ensure stability, transmitting boundary conditions and $B_\phi$ reconnection was adopted \citep[see][for details]{Cho_etal_24}. 
For both runs, a spherical static mesh grid (modified Kerr-Schild) was used with a resolution of 288x128x128.

\begin{figure*}
    \centering
    \includegraphics[width=0.495\linewidth]{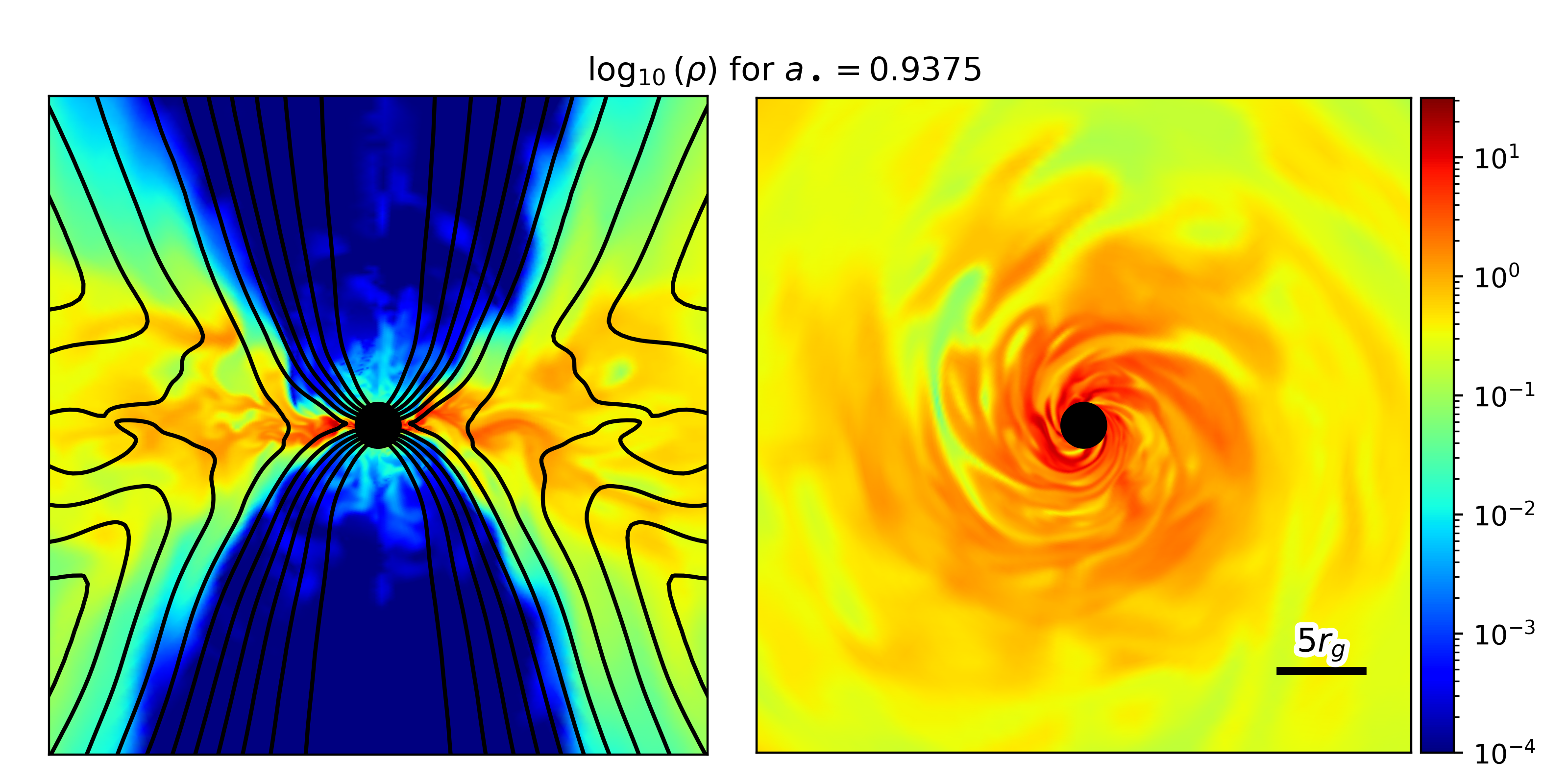}
    \includegraphics[width=0.495\linewidth]{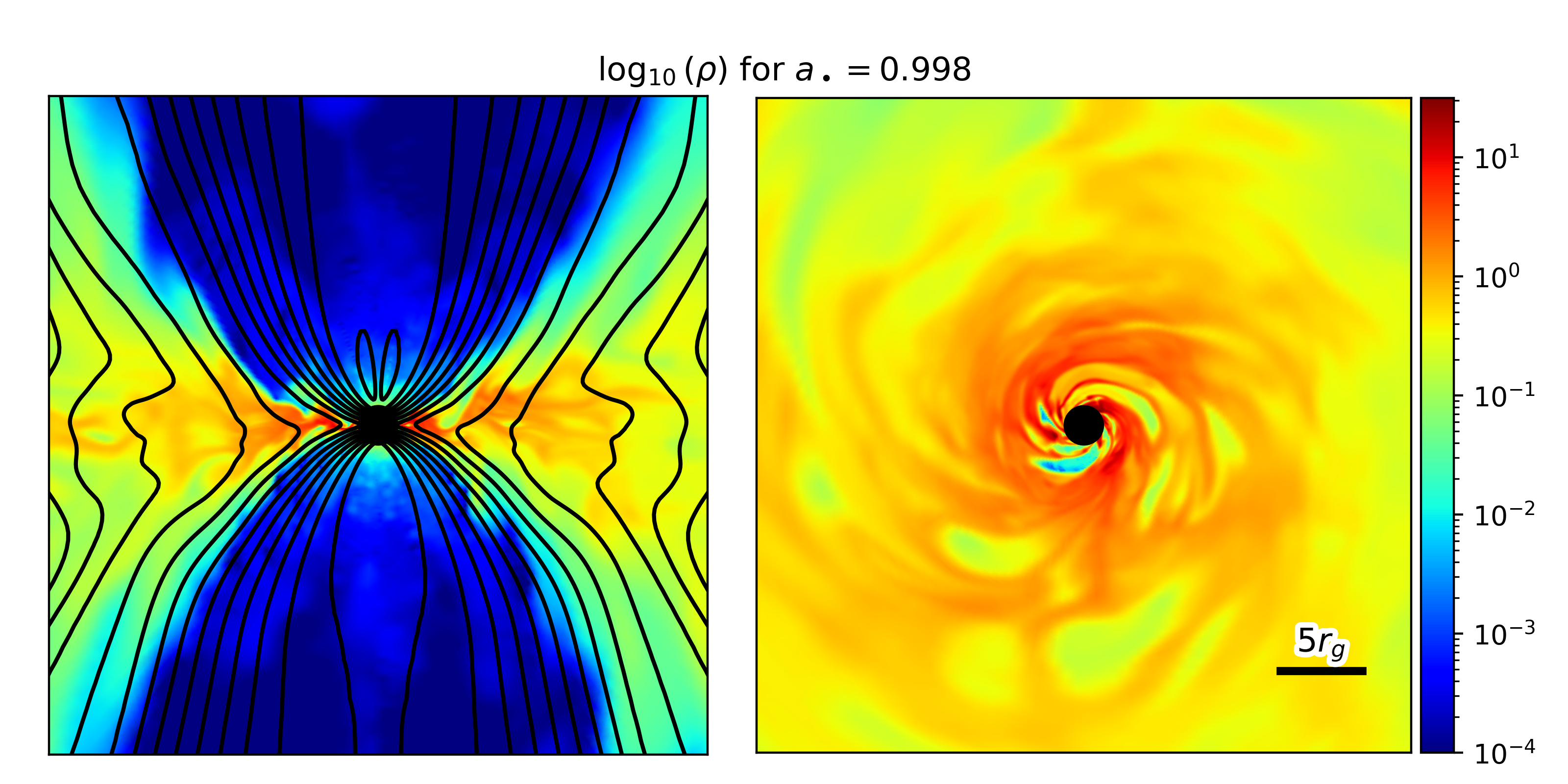}
    \caption{Above are vertical and midplane slices of the GRMHD snapshot from both the $a_\bullet=0.9375$ (left) run and the $a_\bullet=0.998$ (right) run. The color denotes the log density and the contour lines show the poloidal magnetic field.}
    \label{fig:exampleKharma}
\end{figure*}

\subsection{GRRT Imaging}
We use the General Relativistic Radiative Transfer (GRRT) code {\sc IPOLE} to produce full-Stokes images from GRMHD snapshots \citep{Moscibrodzka_Gammie_18}.  A null geodesic is computed for each pixel, along which the full set of Stokes parameters, considering synchrotron emission, absorption, and Faraday effects.  We assume only thermal electron distribution functions and zero emission from regions with $\sigma \equiv b^2/\rho>1$.  All images are computed at 230 GHz assuming the ``fast-light'' approximation.  We consider both primary EHT targets, \sgra and \m87, for which we adopt $M=4.14\times 10^6M_\odot$, $d=8.127$ kpc, \& $F=2.4$ Jy \citep[for consistency with][]{EHT_5_22}, and $M=6.2\times 10^9M_\odot$, $d=16.9$ Mpc, \& $F=0.6$ Jy \citep[for consistency with][]{EHT_M87_25} respectively. For each object, we rescale the mass density, magnetic field strength, and internal energy to achieve the appropriate flux density \citep[see e.g.,][]{Wong_etal_22}.

Within our models, ions and electrons are not in thermal equilibrium and instead we define their temperature ratio for two regimes: 1.) the region where gas pressure dominates and 2.) the region where magnetic pressure dominates \citep{Moscibrodzka_Falcke_Shiokawa_16}. The first regime is controlled by the parameter $R_{high}$ and the second by $R_{low}$. The equation for these two parameters is written below, where $T_p$ is the ion temperature, $T_e$ is the electron temperature, $\beta=P_{gas}/P_{mag}$, and $\beta_{crit}$ is the critical pressure ratio which we take to be 1 \citep{Moscibrodzka_Falcke_Shiokawa_16}. 
\begin{equation}
    \frac{T_p}{T_e}=R_{high}\frac{(\beta^2/\beta_{crit}^2)}{1+(\beta^2/\beta_{crit}^2)}+R_{low}\frac{1}{1+(\beta^2/\beta_{crit}^2)}
\end{equation}
$R_{high}$ and $R_{low}$ are important parameters when imaging because the temperature of ions vs electrons influences synchrotron radiation and self-absorption, which in turn significantly impacts observed flux and Faraday rotation depth \citep{EHT_8_24}. For the \sgra models we use an $R_{high}=160$ and $R_{low}=1$, meanwhile for \m87 models we use $R_{high}=40$ and $R_{low}=1$, each of which are combinations that reasonably reproduce most full-Stokes observables of each target \citep{EHT_5_22,EHT_9_23}. We explore inclinations of $150^\circ$ and $90^\circ$ for the \sgra models and $163^\circ$ and $90^\circ$ for the \m87 models.

\section{Results} \label{results}
\subsection{GRMHD Properties}
Snapshots of the $a_*=0.9375$ and $a_*=0.998$ simulations are shown in figure \ref{fig:exampleKharma}. Modest floor artifacts are visible in the polar regions, motivating our choice to zero the density in these regions during ray-tracing. The scale heights of both disks within the plunging region are comparable and turbulence can be seen within both flows.
\subsubsection{Accretion Rate}
The accretion rate is calculated at $5r_g$ as
\begin{equation}
    \dot{M}_\bullet = -\iint \left( \rho u^r\sqrt{-g}\right)_{r=5r_g} d\theta d\phi.
    \label{eqn:mdot}
\end{equation}

\begin{figure*}
    \centering
    \includegraphics[width=\linewidth]{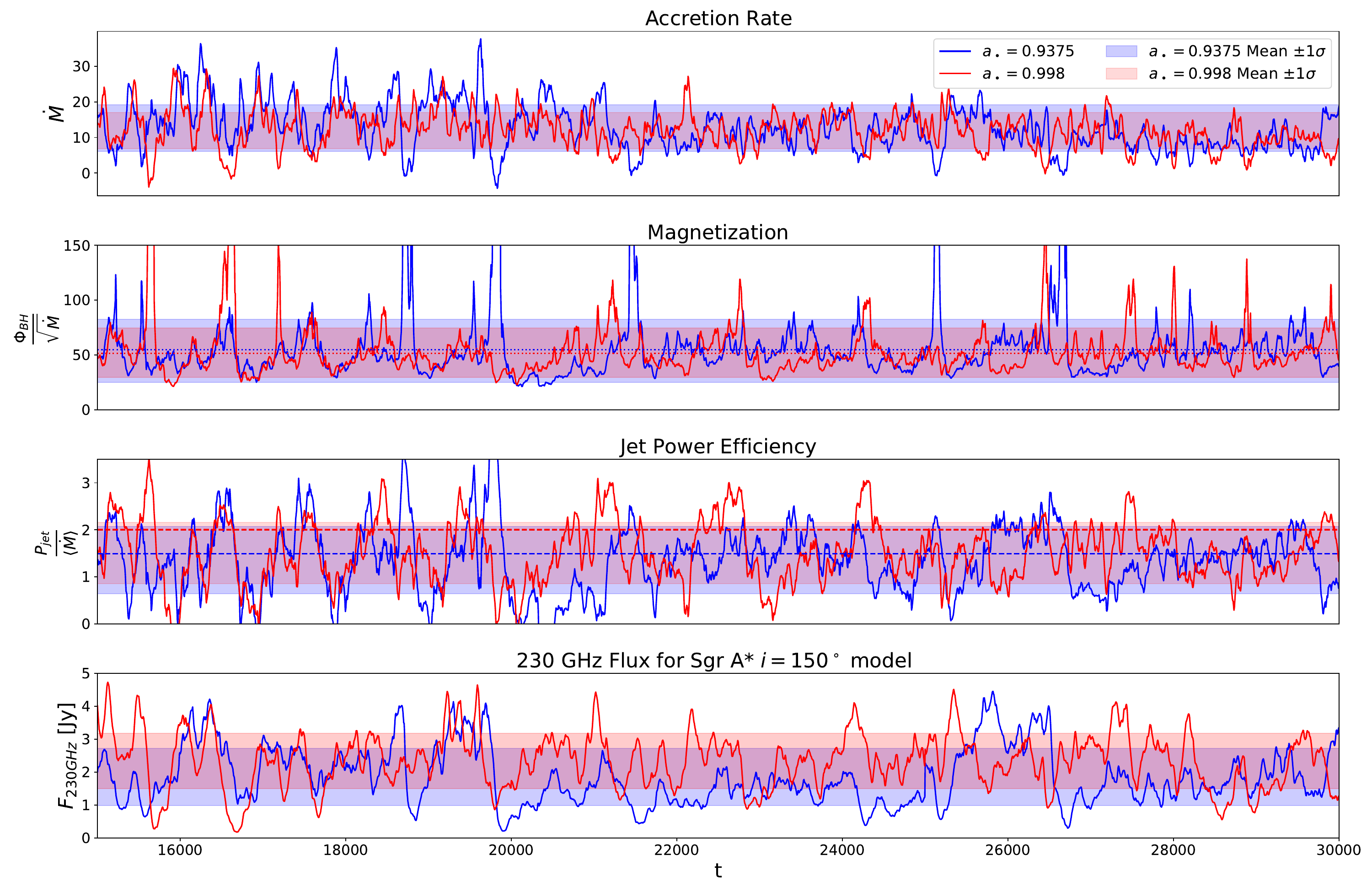}
    \caption{Above plots demonstrate the accretion rate, magnetization, jet power efficiency, and 230GHz flux as a function of time for both $a_\bullet=0.9375$ (blue) and $a_\bullet=0.998$ (red). The highlighted blue and red regions show the mean values $\pm 1\sigma$ (where $\sigma$ denotes a standard deviation) for $a_\bullet=0.9375$ and $0.998$ respectively. The dashed lines seen in the magnetization and jet power efficiency graphs demonstrate the expected mean values based on \cite{Narayan_etal_22} and \cite{Tchekhovskoy_Narayan_McKinney_10}, where magnetization values are $\sim$ 54.735 and $\sim$ 51.612 for $a_\bullet=0.9375$ and $a_\bullet=0.998$ respectively. The reported 230GHz flux is for our Sgr A* model.}
    \label{fig:accretionRate}
\end{figure*}

In code units, as displayed in figure \ref{fig:accretionRate}, the average accretion rate for $a_\bullet=0.9375$ versus $a_\bullet=0.998$ is $\sim$ 12.665 and $\sim$ 11.942 respectively. While they are not significantly different, we do still note the decrease in accretion rate for the near-maximal case. 

In order to estimate the variability of various properties presented below we calculate $\sigma/\mu$, where $\mu$ is the mean value and $\sigma$ is the standard deviation \citep{Chan_etal_2015}.
We measure $\sigma/\mu$ in $530 \ t_g$ intervals, then report the mean among these intervals.  
The $\sigma/\mu$ of $\dot{M}_\bullet$ for the $a_\bullet=0.9375$ and $a_\bullet=0.998$ simulations respectively were 0.399 and 0.371.
This indicates that the near-maximal case has less variable accretion rate, which is important as that is expected to also produce lower total intensity variability. Currently, within the EHT collaboration, most models, including the 0.9375 models, overestimate the variability of Sgr A* when compared to observations \citep{EHT_5_22}.  The lower variability in the $a_\bullet=0.998$ is helpful, but not enough to explain this discrepancy.

\subsubsection{Magnetic Field and Jet Power}
As plasma accretes onto the black hole, frozen-in magnetic field lines thread through the BH horizon. This magnetic flux is most commonly measured via the dimensionless the magnetic flux parameter $\phi_{BH}$, which is calculated using the equation below \citep{Tchekhovskoy_Narayan_McKinney_11, Narayan_etal_22}.
\begin{equation}
    \phi_{BH}(t) = \frac{\sqrt{4\pi}}{2\sqrt{\dot{M_\bullet}(t)}}\int_\theta \int_\phi |B^r\sqrt{-g}|_{r_H}d\theta d\phi
\end{equation}
The $\sqrt{4\pi}$ term converts the magnetic field strength $B^r$ from Heaviside-Lorentz units to Gaussian units, $\dot{M_\bullet}$ is the rest mass inflow rate through the disk as a function of time, $g$ is the determinant of the metric, and $r_H$ is the event horizon radius.  This quantity is often used to characterize the magnetization of the accretion flow, with MADs achieving $\phi \sim 50$ in our units.

Using {\sc PyHARM}, $\phi_{BH}$ is tracked as a function of time and magnetic flux is measured at $r_H$ while accretion rate is measured at $5r_g$. Previous studies of GRMHD simulations found the following equation for $\phi_{BH}$ as a function of spin \citep{Tchekhovskoy_12,Narayan_etal_22}. 
\begin{equation}
    \label{eqn:TchekovskoyMagEqn}
    \phi_{BH}(a_\bullet) = -20.2a_\bullet^3 - 14.9a_\bullet^2 + 34a_\bullet + 52.6
\end{equation}

The average $\phi_{BH}$ for $a_\bullet=0.9375$ versus $a_\bullet=0.998$ is $\sim$ 53.756 and $\sim$ 52.143 respectively which is relatively close to values $\sim$ 54.735 and $\sim$ 51.612 predicted by \autoref{eqn:TchekovskoyMagEqn}. Similar to the accretion rate, we find lower variability in $\phi$ for the higher spin case: $\sigma/\mu=0.376$ for $a_\bullet=0.9375$ and $\sigma/\mu=0.314$ for $a_\bullet=0.998$. This variability, however, appears to be dominated by variability in the accretion rate (in the denominator of $\phi$).

In MAD simulations with large spin, the magnetic field strength is directly tied to the jet power $P_\mathrm{jet}$ via the \citet{Blandford_Znajek_77} mechanism.  Here, a spinning BH twists the magnetic field threading the black hole into coils, forming a large toroidal field component that powers the jet via $P_\mathrm{jet} \propto \dot{M}_\bullet a_\bullet^2\phi^2$. \citep{Blandford_Znajek_77,Davis_Tchekhovskoy_20}. In our models, we measure the jet power at $5r_g$, where it is defined as the outgoing energy flux ($\dot{E}$) without any contribution from the rest-mass energy.

\begin{minipage}{.5\linewidth}
\begin{equation}
\dot{E} = \iint T^r_t \sqrt{-g} d\theta d\phi
\end{equation}
\end{minipage}%
\begin{minipage}{.5\linewidth}
\begin{equation}
P_\mathrm{jet} = \dot{M_\bullet}c^2 - \dot{E}c^2
\end{equation}
\end{minipage}

For MAD simulations, \cite{Tchekhovskoy_Narayan_McKinney_10} found that the jet efficiency $\eta\equiv P_\mathrm{jet}/\langle\dot{M}_\bullet c^2\rangle$ followed the the BZ power for moderate spins, but preferred corrections as $a_\bullet \to 1$ via 
\begin{equation}
    \eta_{BZ6} =\frac{\kappa}{4\pi} \phi_{BH}^2 \Omega_{H}^2[1+1.38\Omega_{H}^2-9.2\Omega_H^4],
    \label{eqn:jetpowerEffPredict}
\end{equation}
where $\Omega_H=\frac{a_\bullet}{2r_H}$ is the horizon angular velocity and $\kappa=0.05$ is a constant whose precise value is determined by the jet's field geometry. It is of interest to us whether these higher order corrections are preferred in our simulations as well.

\begin{figure}
    \centering
    \includegraphics[width=0.99\linewidth]{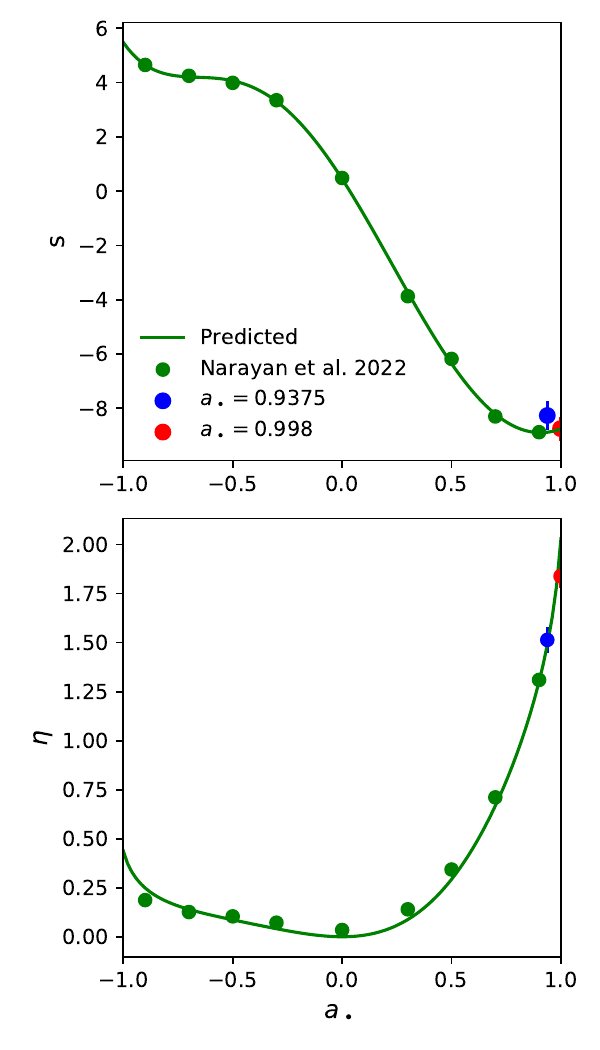}
    \caption{The spin-up parameter (top) and jet power efficiency (bottom) as a function of spin. Green indicates previously calculated values and the predicted curve from \cite{Narayan_etal_22}. The blue and red dots indicate our observed average jet power efficiency. The error bars indicate $\pm5\sigma_s$ based on our calculation of $\sigma_s$, which is shown in equation \ref{eqn:sUncertainty}, and $\pm5\sigma_\eta$ which is calculated similarly. (Note that $\sigma_s$ and $\sigma_\eta$ are distinct from the standard deviation shown in figure \ref{fig:accretionRate}.)}
    \label{fig:spinup_n_eta}
\end{figure}

We compare jet power efficiencies for both spin cases to one another and to the expected value given by equation \ref{eqn:jetpowerEffPredict}. 
Our results find that the measured jet power efficiency is, on average, in very good agreement with the values predicted by \autoref{eqn:jetpowerEffPredict} for both the $a_\bullet=0.9375$ and $a_\bullet=0.998$ case, as seen in figure \ref{fig:spinup_n_eta}. In particular, for the near-maximal case, the average efficiency is $\approx 1.839$ meanwhile the predicted efficiency was $\approx2.0$, a difference of only $\sim8.75\%$. For comparison, $\eta$ for $a_\bullet=0.9375$ was $\approx 1.347$ while the predicted value was $\approx 1.488$, giving a difference of $\sim10.47\%$. For both cases, omitting the terms with $\Omega_H^4$ and $\Omega_H^6$ in \autoref{eqn:jetpowerEffPredict} worsens the agreement. 

When $\eta$ is greater than unity, the jet is extracting more energy and angular momentum than is gained by the disk, leading to a black hole spin down \citep{Narayan_etal_22}. The spin-up of a black hole is analyzed using the dimensionless spin up parameter $s$ as defined below where a negative value indicates spin down \citep{Gammie_04,Shapiro_05}. 

\begin{equation}
    s = \frac{d(J/M^2)}{dt}\frac{M_\bullet}{\dot{M_\bullet}}=\frac{da_\bullet}{dt}\frac{M_\bullet}{\dot{M_\bullet}}
    \label{eqn:spinup}
\end{equation}

We approximate the standard error of the mean spin-up parameter based on the equation \ref{eqn:sUncertainty}, where $\dot{J}$ is the angular momentum flux and $\dot{E}$ is the aforementioned energy flux which relate to $s$ via equation \ref{eqn:alts}. N is the number of samples taken, which in this case is 3,001. $\sigma_{\dot{J}}$, $\sigma_{\dot{E}}$, and $\sigma_{\dot{M}}$ are the standard deviations for each variable. 

\noindent\begin{minipage}{.5\linewidth}
\begin{equation}
\dot{J} = -\iint T^r_\phi \sqrt{-g} d\theta d\phi
\end{equation}
\end{minipage}%
\begin{minipage}{.5\linewidth}
\begin{equation}
s = \frac{\dot{J}}{\dot{M_\bullet}} - 2a_\bullet\frac{\dot{E}}{\dot{M_\bullet}}
    \label{eqn:alts}
\end{equation}
\end{minipage}
\begin{equation}
    \sigma_s = \frac{1}{\sqrt{N}}\sqrt{\left(\frac{\partial s}{\partial \dot{J}}\sigma_{\dot{J}}\right)^2 + \left(\frac{\partial s}{\partial \dot{E}}\sigma_{\dot{E}}\right)^2+\left(\frac{\partial s}{\partial \dot{M_\bullet}}\sigma_{\dot{M_\bullet}}\right)^2}
    \label{eqn:sUncertainty}
\end{equation}

\noindent\begin{minipage}{.5\linewidth}
\begin{equation}
\frac{\partial s}{\partial \dot{J}}=\frac{1}{\langle \dot{M_\bullet} \rangle}
\end{equation}
\end{minipage}%
\begin{minipage}{.5\linewidth}
\begin{equation}
\frac{\partial s}{\partial \dot{E}}=\frac{2a_\bullet}{\langle \dot{M_\bullet} \rangle}
\end{equation}
\end{minipage}

\begin{equation}
    \frac{\partial s}{\partial \dot{M_\bullet}}=\frac{\langle J \rangle}{\langle \dot{M_\bullet} \rangle^2} + \frac{2a_\bullet\langle \dot{E} \rangle}{\langle \dot{M_\bullet} \rangle^2}
\end{equation}

As seen in figure \ref{fig:spinup_n_eta}, the predicted spin-up parameters agree relatively well for both spins. The $a_\bullet=0.9375$ case shows slightly less negative results, indicating less rapid spin-down of the black hole than expected. 

Jet structure may also be examined by identifying the $\sigma=1$ boundary. We solve for $k_0$ and $k$ in the equation

\begin{equation}
    \log_{10}x = k_0 + k\log_{10}z
    \label{eqn:jetStructure}
\end{equation}

\noindent by first azimuthally averaging $\sigma(x,z)$, selecting cells within $0.9<\sigma<1.1$, and minimizing the least-squares metric as evaluated by differencing the prediction of \autoref{eqn:jetStructure} with the true values of $\sigma$.  When doing so, we weight each point by $\frac{1}{|1-\sigma|}$, chosen to apply larger weight to values with $\sigma$ closer to 1.

Roughly speaking, the inner curve of the jet becomes shallower as $k_0$ increases and the large-scale jet width goes as $x\propto z^k$.
We find average $k_0$ values of $0.671$ and $0.728$ for $a_\bullet=0.9375$ and $a_\bullet=0.998$ respectively, as well as $k$ values of $0.479$ and $0.429$. These $k_0$ values are lower than expected given the reported values in \citet{Narayan_etal_22}, which indicates that the $\sigma=1$ region extends to a slightly larger region in the mid-plane, however the $k$ measurements are in greater agreement. These values are not statistically significantly different, however it is notable that the higher spin model had a lower $k$ value, which contradicts the expected increase in $k$ with increased spin. 

\subsection{GRRT Properties}
In order to roughly approximate the EHT's observational capabilities we apply a Gaussian blur to images produced such that they have a final resolution of $20 \ \mu$as, unless otherwise specified.

\subsubsection{Total Intensity Images}
Using {\sc IPOLE} we created movies for the $a_\bullet=0.998$ and $a_\bullet=0.9375$ runs over the course of $15,000-30,000t_g$. 
The videos aren't directly comparable frame by frame since the variability is driven by random instantiations of the turbulence. That being said we are able to evaluate trends over time by evaluating the light curve shown in figure \ref{fig:accretionRate}.

The light curves over all the models between the two spins exhibit similar variability amplitudes and timescales.  Variability is of particular importance when modeling Sgr A* because, as noted in \cite{EHT_5_22}, nearly all models of Sgr A* exhibit higher variability than the observed $\sigma/\mu\sim0.1$\citet{Wielgus_etal_22}. Based off of our results, $a_\bullet=0.998$ models wouldn't solve this problem, as for the $i=150^\circ$ model the variability is comparable between our $a_\bullet=0.998$ run ($\sigma/\mu\approx0.287$) and our $a_\bullet=0.9375$ run ($\sigma/\mu \approx0.277$). 

Aside from the light curve, we can also examine the overall image differences by taking an average over the full time span.
\begin{figure*}
    \centering
    \includegraphics[width=0.95\linewidth]{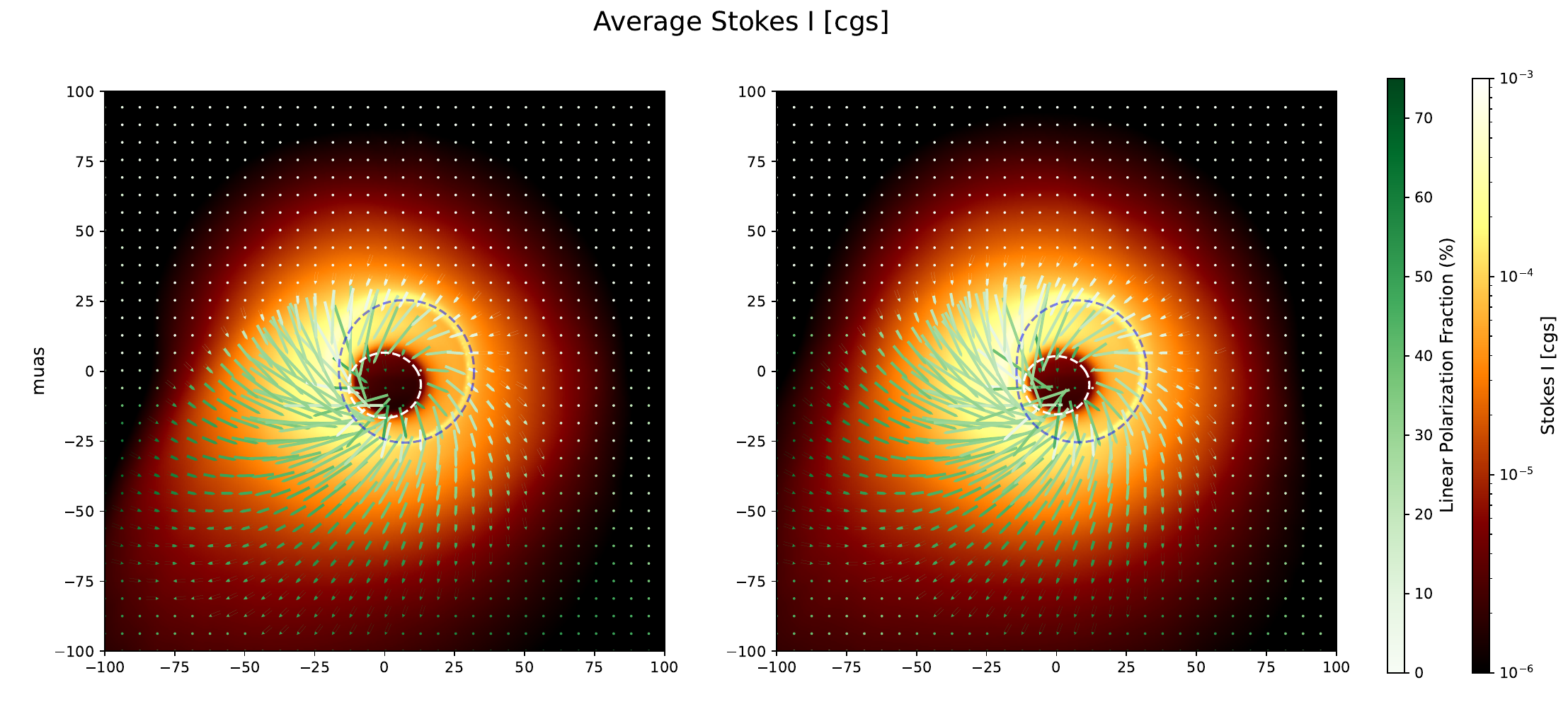}
    \caption{Above are the time averaged images for the $a_\bullet=0.9375$ (left) and $a_\bullet=0.998$ (right) \sgra runs  for a resolution of 0.625$\mu$as and inclination of $150^\circ$. The average is taken over the course of $15,000-30,000t_g$ and involves averaging all stokes parameters. The colorplot shows the average Stokes I in cgs units. The short lines show the average linear polarization direction and the color reflects the linear polarization fraction with darker green indicating a higher fraction. The navy dashed line indicates the critical curve and the white dashed line indicates the approximate lensed horizon \citep{Chael_Johnson_Lupsasca_21}.}
    \label{fig:averageIpole}
\end{figure*}
The average images for the $a_\bullet=0.9375$ and $a_\bullet=0.998$ runs, shown in figure \ref{fig:averageIpole}, are remarkably similar, demonstrating the difficulty of determining spin using such images. 
 
Using the strategy outlined in \cite{Medeiros_etal_22} we may also examine the ring asymmetry for both spin cases. For the $i=150^\circ$ \sgra models we find that the average asymmetry is $1.385 \pm 0.397$ for $a_\bullet=0.9375$ and $1.370\pm0.406$ for $a_\bullet = 0.998$. For the $i=90^\circ$ \sgra models, however, we find an average asymmetry of $1.670 \pm 0.442$ for $a_\bullet=0.9375$ and $1.781\pm0.562$ for $a_\bullet = 0.998$. This agrees with previous findings showing increased asymmetry for more edge-on inclinations, however the different spin values are too similar to yield significantly different results \citep{Medeiros_etal_22}. 

The same trend is seen in \m87 models, although higher asymmetries were reported on average, which is consistent with observations. Additionally, the above values were calculated assuming a resolution of $0.625 \ \mu$as and so images matching current EHT resolution will likely show further diminished differences in asymmetry.  This suggests that while ring asymmetry may be a useful metric to estimate inclination, the dependence is too weak to confidently distinguish near-maximal spins. Distributions of asymmetry measurements for both models can be found in the rightmost panels of figure \ref{fig:obsViolinPlots}.

\subsubsection{Polarization}

\begin{figure*}
    \centering
    \includegraphics[width=0.975\linewidth]{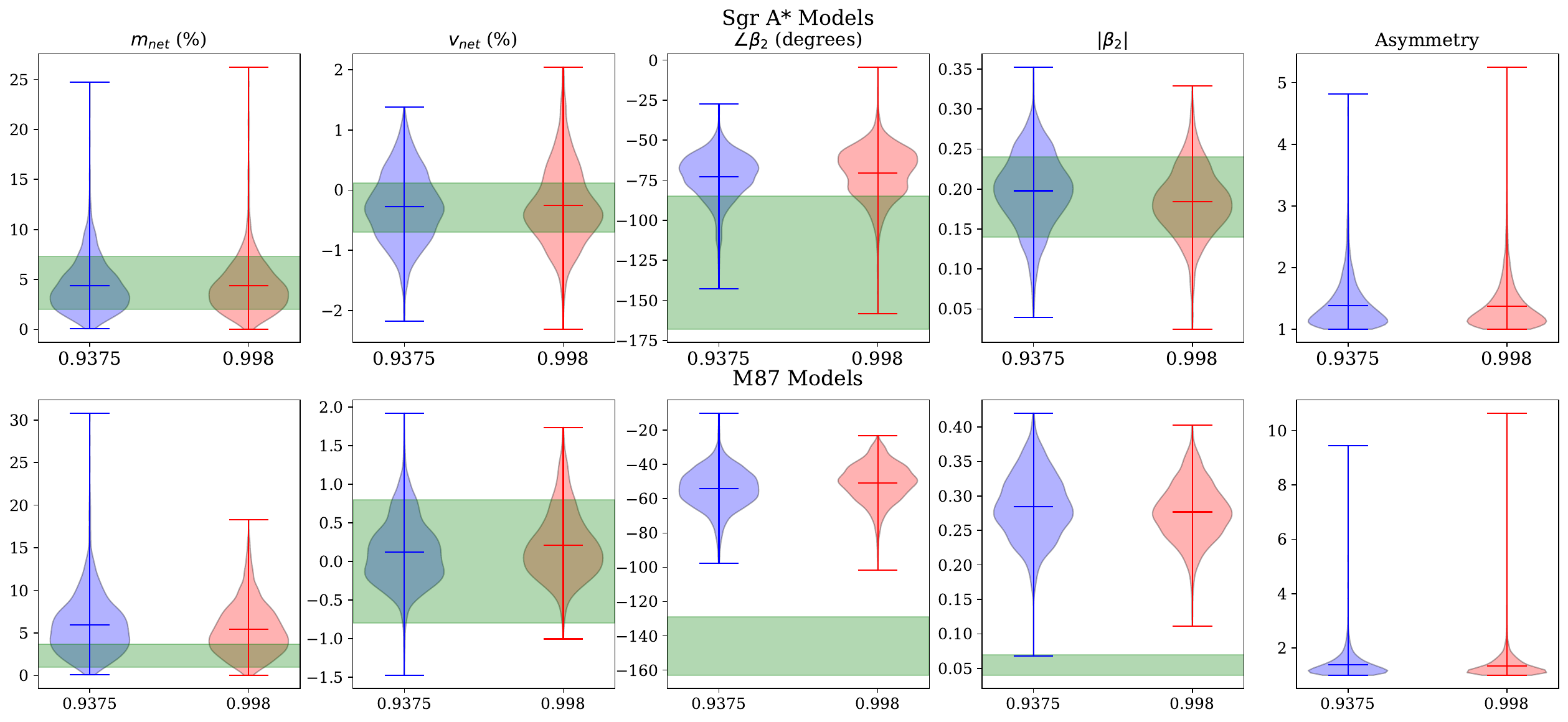}
    \caption{Above are violin plots of observable parameters for the \sgra inclination $150^\circ$ and \m87 inclination $163^\circ$ models compared to observational constraints. Note that the confidence interval for the \sgra models' $\angle \beta_2$ is the derotated range.}
    \label{fig:obsViolinPlots}
\end{figure*}
Within the {\sc IPOLE} images we are also able to examine polarization. This is a key aspect of the images as most of the emission EHT detects comes from synchrotron radiation, which can probe the magnetic field structure near the black hole. 

Below we will discuss the most relevant polarization metrics: $m_{net}$, $v_{net}$, and the phase and magnitude of $\beta_2$.

\paragraph{$m_{net}$}
The spatially unresolved linear polarization fraction, obtainable by e.g., a single-dish measurement of a source.  This is calculated via.
\begin{equation}
    m_{net}=\frac{\sqrt{(\Sigma_{pixel}Q)^2+(\Sigma_{pixel}U)^2}}{\Sigma_{pixel}I}
\end{equation}
where $I, Q, $ and $U$ are the Stokes parameters.

\paragraph{$v_{net}$}
The analogous quantity for circular polarization, calculated via
\begin{equation}
    v_{net}=\frac{\Sigma_{pixel}V}{\Sigma_{pixel}I}
\end{equation}

Small amounts of circular polarization naturally occur as part of synchrotron radiation, however the primary source of circular polarization within EHT images would be due to Faraday conversion \citep{Ricarte_Qui_Narayan_21,EHT_9_23}. 
$v_{net}$ specifically helps inform the direction of the poloidal field in the observer's line of sight \citep{EHT_8_24}.

\paragraph{$\beta_2$} \label{b2}
The 2nd mode decomposition coefficient defined by the equation below \citep{Palumbo_Wong_Prather_20}. 
\begin{equation}
    \beta_2 = \frac{1}{I_{ann}}\int_{\rho_{min}}^{\rho_{max}}\int_0^{2\pi}P(\rho, \varphi)e^{-i2\varphi} \rho d\varphi d\rho
\end{equation}
where $I_{ann}$ is the total Stokes $I$ flux in the annulus, $\rho_{min}$ and $\rho_{max}$ are the radial extent of the annulus, and $P(\rho, \varphi)$ is the complex valued polarization field $Q(\rho,\varphi)+iU(\rho,\varphi)$.

In a simpler sense, $\beta_2$ is a complex number that summarizes the rotationally-symmetric structure of linear polarization ticks. The phase encodes the pitch angle, while the magnitude encodes the strength of this mode\citep{Palumbo_Wong_Prather_20}. 

Because the underlying magnetic field structure is rotationally symmetric, numerous studies have demonstrated that $\beta_2$ is a sensitive tracer of spin \citep{Palumbo_Wong_Prather_20, Emami_etal_23, Qiu_etal_23, Chael_etal_23}.  Thus, it is of interest how rapidly this observable changes between our two spin values.

\paragraph{Polarimetric Results}

Our two high spin simulations demonstrate remarkably similar values to one another for the noted polarimetric quantities for both \sgra and \m87, shown in top and bottom rows of Figure~\ref{fig:obsViolinPlots} respectively.

Both spin values seem to fit well with \sgra constraints for all metrics, although the agreement with $\angle \beta_2$ is marginal. This is fully consistent with previous studies \citep{EHT_8_24}.  We report no significant difference in these polarized metrics between $a_\bullet=0.9375$ and $a_\bullet=0.998$, suggesting that EHT observations of Sgr A$^*$ are at present are consistent with both values of BH spins.

For \m87, however, all of the metrics outside of $v_{net}$ don't fit with observational constraints. This is expected as lower spin and retrograde models tend to be favored for \m87, especially when using {\sc KHARMA} \citep{EHT_2_25}.

\subsubsection{The Photon Ring}

Given the remarkable similarity of the fluid and image properties of these two simulations, we turn to the photon ring as a spin discriminant.  The photon ring is comprised of marginally bound photons that complete at least one half orbit around the black hole, resulting in a sharp feature whose size and shape encode the black hole mass and spin.  Spatially resolving this ring is one of the key scientific goals of BHEX \citep{Johnson_etal_24}. For that reason, it's important to check the observability in differences of the photon ring between $a_\bullet=0.998$ and $a_\bullet=0.9375$. 

\begin{figure*}
\centering
\includegraphics[width=.95\linewidth]{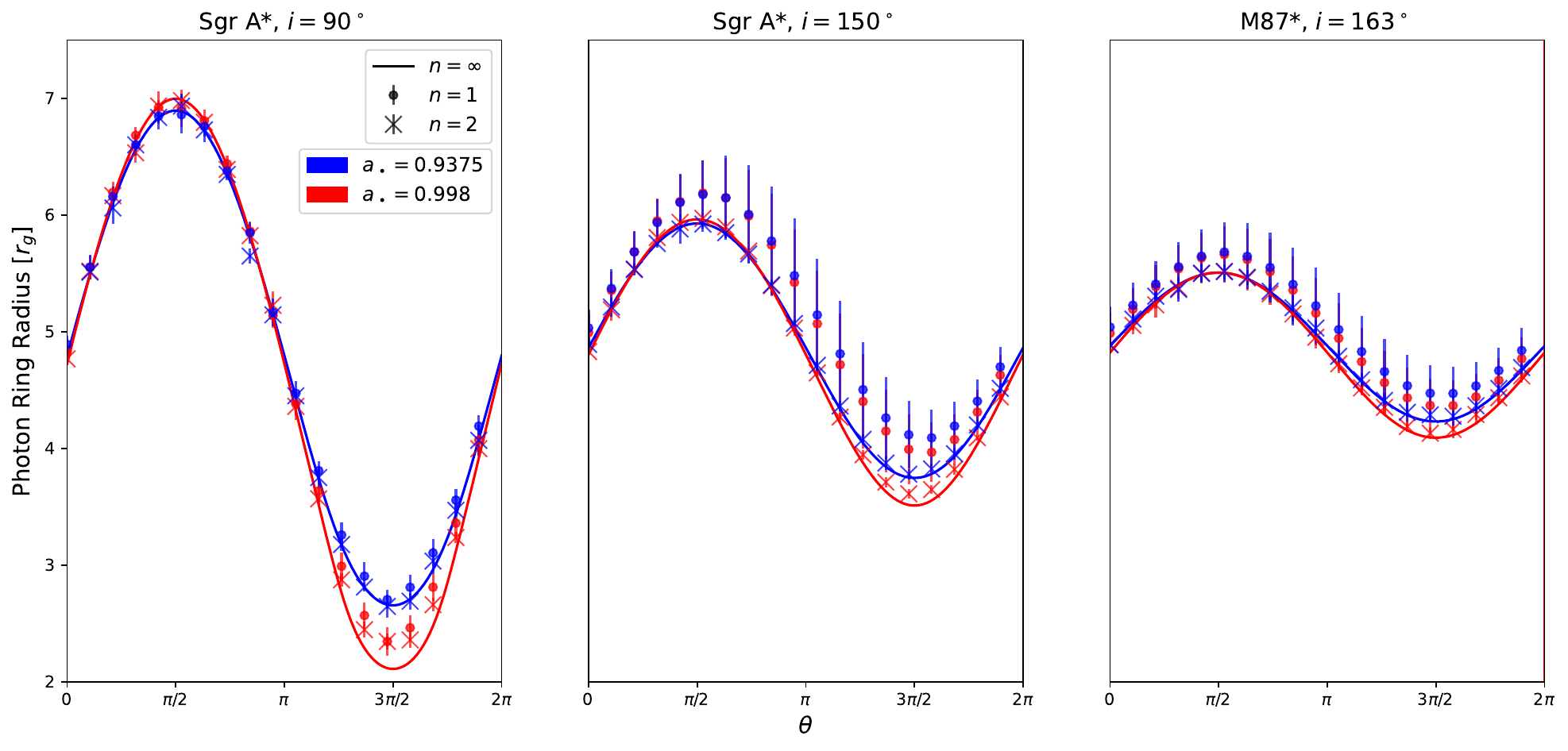} 
\caption{Average radius of the photon ring measured from the center of the image for both $a_\bullet=0.9375$ (blue) and $a_\bullet=0.998$ (red).  Bardeen coordinates are adopted.  The analytically calculated values for the expected $n=\infty$ rings are also plotted for reference. Error bars denote the average ring width. n=1 photon rings were imaged with a 0.625 $\mu$as resolution and n=2 photon rings were imaged with a 0.3125 $\mu$as resolution.}
\label{fig:photonring}
\end{figure*}

In \autoref{fig:photonring}, we plot photon ring radii and widths as markers with error bars, compared with analytic curves as a function of angle in the image.  Photon ring ``sub-rings'' are indexed by the number $n$ of half-orbits the photon completes before escaping to infinity.  Analytic expressions exist for the $n=\infty$ ring, or ``critical curve,'' towards which these rings converge as $n$ increases.  For finite $n$, the exact shape retains some sensitivity to the underlying emission structure, although as demonstrated here, the deviation is small.

The most intense effects to the critical curve, and therefore the photon ring, occurs when the black hole is viewed at an inclination of $90^\circ$. For this reason, we image the $a_\bullet=0.998$ and $0.9375$ photon rings at an inclination of $90^\circ$ despite the fact that such inclinations aren't favored for \sgra or \m87. {\sc IPOLE} isolates photon rings of different $n$ by keeping track of turning points in the $\theta$ coordinate as the geodesic evolves.  For an inclination of $90^\circ$, this convention is less appropriate and our sub-images labeled $n$ are actually calculated using $n+1$ in {\sc IPOLE}.

As seen in figure \ref{fig:photonring}, the photon ring radius notably changes between the $a_\bullet=0.998$ and $a_\bullet=0.9375$ cases. As expected, the $n=2$ images exhibit  a smaller offset from the $n=\infty$ curve than $n=1$ by a factor of $\sim1.537$ for the \sgra $i=150^\circ$ case.  In practice, the largest difference between the two spin values occurs on the Doppler beamed side of the image, where the photon ring is flatter for larger $a_\bullet$ (see \autoref{fig:averageIpole}).  However, this distortion of the n=2 photon ring is only approximately $1.88 \mu as$ for the edge-on \sgra case, and more detailed synthetic data tests will be necessary to determine if the difference can be detected at BHEX resolution.  

In practice, photon ring detection will likely proceed by jointly fitting a constrained ring shape predicted by GR superposed with a flexible image.  This process has already been performed for existing EHT data, and additional data from BHEX will significantly improve constraints \citep{Broderick_etal_22b}.  In performing this process, our images demonstrate that additional calibration is necessary to accommodate the angle-dependent bias away from the $n=\infty$ curve exhibited by our $n=1$ images, as is currently performed on the $n=0$ image \citep{EHT_6_19,EHTC_6_22}.

\section{Discussion \& Conclusions} \label{conclusion}

In this work, we perform two GRMHD simulations of MAD accretion flows, one with spin $a_\bullet=0.9375$ and another with spin $a_\bullet=0.998$, to quantify evolution in emergent fluid properties as $a_\bullet \to 1$.  Our results can be summarized as follows:

\begin{itemize}
    \item The fluid properties of the two simulations are remarkably similar, each attaining levels of magnetization, jet efficiency, and variability that are indistinguishable given the scatter of these models.
    \item Characteristics of the full-Stokes images of the two simulations are also very similar, ranging from the asymmetry in total intensity, the linear polarization pattern, and the circular polarization fraction.
    \item Given the similarity of the fluid solutions that develop, we find that the only clear discriminant between the two models is through the photon ring, whose size and shape can be constrained by the Black Hole Explorer (BHEX) for Sagittarius A* and M87*. 
\end{itemize}

\noindent \citet{Gelles_etal_25,Gelles_etal_26} propose a way to constrain spin by resolving a polarization flip along the jet associated with the light cylinder.  Unfortunately, the radius of the light cylinder evolves with spin via $R_L \sim 1/a_\bullet$, making this signature more discriminating at low spin values than high ones.

In conclusion, we believe that models with $a_\bullet=0.9375$ can be taken to represent models with spin values up to $a_\bullet=0.998$ for most practical purposes.  In particular, the best-bet model of Sagittarius A* following polarimetric analysis had $a_\bullet=0.9375$ \citep{EHT_8_24}; our results imply that models with larger spin values could be equally accommodated.  We do not find any way to observationally distinguish these spin values apart from photon ring measurements, as would be enabled by the Black Hole Explorer (BHEX).  For performing such measurements, the calculations performed in this work will enable us to calibrate offsets from the ``critical curve'' shown in \autoref{fig:photonring} as $a_\bullet \to 0.998$.

\begin{acknowledgments}
We thank Yajie Yuan for very helpful feedback and Zachary Gelles for helpful comments regarding constraining spin with jets. T.A.T. was supported by the NSBP/SAO EHT Scholars Program, the Gordon and Betty Moore Foundation (Grant \#10423), and generous support from Mr.\ Michael Tuteur and Amy Tuteur, MD. A.R., C.P., and H.C. were partially supported by the Black Hole Initiative at Harvard University, which is funded in part by the Gordon and Betty Moore Foundation (Grants \#13526 \#8273). It was also made possible through the support of a grant from the John Templeton Foundation (Grant \#63445 \#62286). The opinions expressed in this publication are those of the author(s) and do not necessarily reflect the views of these Foundations. We also acknowledge financial support from the National Science Foundation (AST-2307887).
\end{acknowledgments}

\bibliography{sample7}{}
\bibliographystyle{aasjournalv7}

\end{document}